\documentclass[aps,pra,twocolumn]{revtex4}
\RequirePackage{epsfig}
\def\vec#1{{\bf #1}}
\def\op#1{\hat{#1}}
\def\ket#1{| #1 \rangle}
\def\bra#1{\langle #1 |}
\def\ave#1{\langle #1 \rangle}

\def\trace#1{\mathop{\rm Tr}\nolimits \left( #1 \right)} 
\def\span{\mbox{span}}
\newtheorem{theorem}{Theorem}

\begin{document}
\bibliographystyle{prsty}
\title{Limits of control for quantum systems: kinematical bounds on the 
       optimization of observables and the question of dynamical realizability}
\author{S.~G.\ Schirmer}
\affiliation{Quantum Processes Group, The Open University, Milton-Keynes, MK7 6AA, United Kingdom}
\author{J.~V.\ Leahy}
\affiliation{Department of Mathematics and Institute of Theoretical Science, University of Oregon, Eugene, OR 97403}
\date{September 28, 2000} 
\begin{abstract}
In this paper we investigate the limits of control for mixed-state quantum systems.  
The constraint of unitary evolution for non-dissipative quantum systems imposes 
kinematical bounds on the optimization of arbitrary observables.  We summarize our 
previous results on kinematical bounds and show that these bounds are dynamically 
realizable for completely controllable systems.  Moreover, we establish improved 
bounds for certain partially controllable systems.  Finally, the question of dynamical
realizability of the bounds for arbitary partially controllable systems is shown to 
depend on the accessible sets of the associated control system on the unitary group 
$U(N)$ and the results of a few control computations are discussed briefly. 
\end{abstract}
\maketitle
\section{Introduction}
\label{sec:Intro}
Recent advances in laser technology have opened up new possibilities for laser 
control of quantum phenomena such as control of molecular quantum states, chemical 
reaction dynamics or quantum computers to mention only a few.  The limited success 
of initially advocated control schemes based largely on physical intuition in both
theory and experiment \cite{00RVMK}, has prompted researchers in recent years to 
study these systems systematically using control theory.  One of the many interesting
questions that has arisen is the issue of dynamical realizability of bounds on the 
optimization of observables imposed by kinematical contraints, which we shall address
in this paper.

The answer to this question is not just of theoretical interest, it is a matter 
of practical importance as well since most algorithms designed to find optimal 
controls \cite{00SGL,98ZR} to drive a system are based on a set of differential 
equations providing necessary but not sufficient conditions for a global maximum, 
i.e., these algorithms may produce controls that steer the system to a local maximum
or minimum, for instance.  Independent knowledge of dynamically attainable bounds 
on the expectation value of the observable makes it possible to determine the 
effectiveness of a given control in achieving the control objective of maximizing 
or minimizing the expectation value of a given observable.
\section{Quantum Statistical Mechanics Model}
\label{sec:MathSetup}
We consider a quantum-mechanical system whose pure states form a separable Hilbert 
space $\cal H$.  A mixed state is an ensemble of orthonormal pure quantum states $\Psi_k$
with a discrete probability distribution assigning each pure state a certain 
probability $w_k$ such that $0 \le w_k \le 1$ and $\sum_k w_k=1.$  Such a state can
always be represented by a density operator $\op{\rho}$ on $\cal H$ with eigenvalue 
decomposition
\begin{equation} \label{Eq:rho}
   \op{\rho} = \sum_k w_k \ket{\Psi_k}\bra{\Psi_k},
\end{equation}
where $w_k$ are the eigenvalues, and $\ket{\Psi_k}$ the corresponding normalized 
eigenstates of $\op{\rho}$.  Unless otherwise specified, we shall use the word 
``state'' in the following to refer to a mixed quantum state represented by a 
density operator $\op{\rho}$. 

If the system is Hamiltonian, the time-evolution of $\op{\rho}$ is given by
\begin{equation} \label{Eq:rhoEvol}
   \op{\rho}(t) = \op{U}(t,t_0) \op{\rho}_0 \op{U}(t,t_0)^\dagger, 
\end{equation}
where $\op{\rho}_0=\op{\rho}(t_0)$ and $\op{U}(t,t_0)$ is the time-evolution 
operator of the system satisfying the time-dependent Schr{\"o}dinger equation
\begin{equation} \label{Eq:Schro}
  i\hbar \frac{\partial}{\partial t} \op{U}(t,t_0) =\op{H} \op{U}(t,t_0).
\end{equation}
$\op{H}$ is the total Hamiltonian of the system.  

Observables are represented by Hermitian operators $\op{A}$ on $\cal H$ and we define 
their expectation value to be the \emph{ensemble average}
\begin{equation} \label{Eq:expect}
  \ave{\op{A}(t)} = \trace{\op{A} \op{\rho}(t)}.
\end{equation}

The aim of controlling the system is to maximize the expectation value of a chosen 
observable $\op{A}$ at a certain target time $t_F$ by driving the system using a 
set of optimal control fields $\vec{f}(t)=(f_1(t),\cdots,f_M(t))$.  If the control 
fields are sufficiently weak, it can be assumed that the system is control-linear,
\begin{equation} \label{Eq:ctrl-lin}
   \op{H} = \op{H}_0 + \sum_{m=1}^M f_m(t) \op{H}_m,
\end{equation}
where $\op{H}_0$ is the internal Hamiltonian of the system and $\op{H}_m$, 
$m\ge 1$, is the interaction Hamiltonian for the field $f_m$.
\section{Kinematical Restrictions on the Dynamics}
\label{sec:KB}
If dissipative effects are negligeable then the quantum system is Hamiltonian
and therefore the evolution of the system has to be unitary no matter how the 
system is driven.  This observation has profound implications.  Given an initial
mixed state $\op{\rho}_0$, the only kinematically attainable target states 
$\op{\rho}(t_F)$ are 
\begin{equation} \label{Eq:KinConstraint}
      \op{\rho}(t_F) = \op{U}(t_F,t_0) \op{\rho}_0 \op{U}(t_F,t_0)^\dagger,
\end{equation}
where $\op{U}(t_F,t_0)$ is a unitary transformation.

This kinematical constraint on the dynamical evolution leads to bounds on the
expectation value of arbitrary observables that are completely independent of 
the control functions and thus impose kinematical restrictions on the 
optimization of observables, summarized by the following two theorems 
\cite{98GSLK}:
\begin{theorem}\label{thm:KBI}
Let $\op{A}$ be a Hermitian operator on $\cal H$ with eigenvalue decomposition 
\begin{equation}
    \op{A}=\sum_{i=1}^m a_i \op{I}(a_i),
\end{equation}
where $\op{I}(a_i)$ is the projector onto the eigenspace $E(a_i)$, and let 
$\lambda_1\ge\lambda_2\ge\cdots\ge\lambda_N$ be the eigenvalues $a_i$, counted 
with multiplicity, and ordered in a decreasing sequence.  Then we have
\begin{equation}
   \sum_{k=1}^N \lambda_{N-k+1} w_k
   \le \trace{\op{A}\op{\rho}(t)}
   \le \sum_{k=1}^N \lambda_k w_k,
\end{equation}
provided that the weights $w_k$ are ordered in a decreasing sequence, i.e.,
$w_1 \ge w_2 \ge \ldots \ge w_k \ge \ldots$.
\end{theorem}
\begin{theorem}\label{thm:KBIb}
$\ave{A(t_F)}$ assumes its upper bound if for all $k$ from 1 to $m$
\begin{equation}
  \span_{j=1,\ldots,d(k)} \ket{\Psi_{r(k,j)}(t_F)} = E(a_k),
\end{equation}
and its lower bound if for all $k$ from 1 to $m$
\begin{equation}
  \span_{j=1,\cdots,d(k)} \ket{\Psi_{r(k,j)}(t_F)} = E(a_{N-k+1}),
\end{equation}
where $d(k)=\dim E(a_k)$, i.e., the dimension of the eigenspace belonging
to the eigenvalue $a_k$, and $r(k,j)=d(1)+\cdots+d(k-1)+j.$
\end{theorem}
Thus, to attain the kinematical maximum for an observable whose eigenvalues 
are distinct with multiplicity one, we need to find a unitary transformation 
that simultaneously maps the initial pure state with the largest probability 
$w_1$ onto the eigenspace corresponding to the largest eigenvalue of $\op{A}$, 
the initial state with the second largest probability $w_2$ onto the eigenspace 
corresponding to the second largest eigenvalue of $\op{A}$, and so forth.  If 
$\op{A}$ is a projector onto a subspace of dimension $d$, then attaining the 
kinematical upper bound requires finding a unitary transformation that maps the 
$d$ initial states with the $d$ largest probabilities onto the eigenspace 
corresponding to the eigenvalue one of $\op{A}$.
\section{Dynamical Realizability of the Kinematical Bounds}
\label{sec:dynReal}
The kinematical bounds derived above are clearly dynamically realizable if 
every unitary operator in $U(N)$ is accessible from the identity operator 
$\op{I}$ via a path that satisfies the dynamical law (\ref{Eq:Schro}), i.e.,
if the system is completely mixed-state controllable.

For control-linear systems, complete controllability can easily be verified
numerically.  Note that combining eqs. (\ref{Eq:Schro}) and (\ref{Eq:ctrl-lin}) 
leads to
\begin{equation} \label{Eq:DL}
  i\hbar\frac{\partial}{\partial t} \op{U}(t,t_0)
 =  \op{H}_0 \op{U}(t,t_0) + \sum_{m=1}^M f_m(t) \op{H}_m \op{U}(t,t_0)
\end{equation}
Setting $x(t)=\op{U}(t,t_0)$ and
\begin{equation}
 \vec{X}_m(x(t)) = -\frac{i}{\hbar} \op{H}_m \op{U}(t,t_0), \quad m=0,\cdots,M,
\end{equation}
equation (\ref{Eq:DL}) becomes
\begin{equation} \label{Eq:Lie-Sys}
   \frac{d x}{d t}  = \vec{X}_0(x(t)) + \sum_{m=1}^M f_m(t) \vec{X}_m(x(t)),
\end{equation}
which defines a control system on the Lie group $U(N)$ of a type studied 
by Jurdjevic and Sussmann \cite{72JS}.  From their results, the following
simple algebraic conditions for complete controllability of control-linear, 
non-dissipative quantum systems have been derived\cite{95RSDRP}:
\begin{theorem}{}\label{thm:controllability} 
If the total Hamiltonian is given by (\ref{Eq:ctrl-lin}), where $f_m$ are 
independent bounded measurable control functions, and $\dim{\cal H}=N<\infty$ then
a necessary and sufficient condition for the system to be completely 
controllable is that the Lie sub-algebra $L_0$ of $u(N)$ (the skew-Hermitian
matrices) generated by $\op{H}_0,\cdots,\op{H}_M$ has dimension $N^2$, or 
equivalently, that the ideal $\ell_0$ of $L(U(N))$ generated by $\op{H}_1,
\cdots,\op{H}_M$ has dimension $N^2-1$.
\end{theorem}
If the system is not control-linear, i.e., the Hamiltonian depends in a
nonlinear way on the control functions $f_m$ then there is in general no 
simple algebraic condition to verify controllability. 

For systems that are not completely controllable, dynamical realizability of
a particular kinematical bound depends on the set of unitary transformations
$\op{U}(t_F,t_0) \in U(N)$ that are accessible from the identity $\op{I}\in 
U(N)$.  More precisely, it depends on whether the intersection of the set of
dynamically accessible target states and the set of states for which the 
expectation value of the chosen observable assumes the kinematical bound is 
empty or not.  Since the set of dynamically accessible target states consists
of all density matrices satisfying (\ref{Eq:KinConstraint}), where $U(t_F,t_0)$ 
is a unitary transformation accessible from the identity in $U(N)$ via a path 
that satisfies the equation of motion (\ref{Eq:Schro}), a crucial step towards 
answering the question of dynamical realizability is to determine the 
accessible sets for the associated control system on $U(N)$.
\section{Dynamically Realizable Bounds for Decoupled Systems}
\label{sec:DS}
Among the systems that are obviously not completely controllable are those 
comprised of non-interacting subsystems.  We shall refer to these systems as
decoupled systems.  

Let us first consider a control-linear Hamiltonian system with a single control,
\begin{equation} \label{Eq:Hf}
   \op{H}(f(t)) = \op{H}_0 + f(t) \op{V},
\end{equation}
where $\op{H}_0$ is the internal Hamiltonian of the unperturbed system and 
$\op{V}$ defines the interaction with the control field $f(t)$.  

In this case, the system is decoupled if there exists a basis $\cal B$ for the
Hilbert space $\cal H$ such that $\op{H}_0$ is diagonal and
\begin{equation} \label{Eq:Vtot}
  \op{V} = \op{V}_1 \oplus \op{V}_2 
         \doteq  \left( \begin{array}{c|c} \op{V}_1 & 0 \\\hline 
                                            0 & \op{V}_2 
                  \end{array} \right).
\end{equation}
Let ${\cal H}_1$ and ${\cal H}_2$ be orthogonal subspaces of $\cal H$ such that 
${\cal H}={\cal H}_1\oplus{\cal H}_2$ and each $\op{V}_i$ maps ${\cal H}_i$ to itself, 
\begin{equation}
  {\cal H} = {\cal H}_1 \oplus {\cal H}_2, \qquad \op{V}_i: {\cal H}_i \rightarrow {\cal H}_i, \quad i=1,2.
\end{equation}
It immediately follows that $\op{H}(f(t))$ is block-diagonal,
\begin{equation}
  {\cal H}(f(t)) = \op{H}_1 \oplus \op{H}_2 
           \doteq \left( \begin{array}{c|c} \op{H}_1 & 0 \\hline
                                            0 & \op{H}_2 
                         \end{array}\right)
\end{equation}
and maps ${\cal H}_i$ to itself for $i=1,2$.  Thus, the two subspaces ${\cal H}_1$ and 
${\cal H}_2$ do not interact.  Let ${\cal B}_i$ be the restriction of the basis 
$\cal B$ to the subspace ${\cal H}_i$, $\op{P}_i$ be the projector onto the subspace 
${\cal H}_i$ and let $N_i$ denote the dimension of ${\cal H}_i$.

Given an observable $\op{A}$ on ${\cal H}$, we define the restricted observables 
$\op{A}_i = \op{P}_i \op{A}$ for $i=1,2$.  Note that $\op{A}_i$ is a Hermitian
operator on the subspace ${\cal H}_i$, i.e.,
\begin{equation}
  \op{A}_i=\op{P}_i\op{A} \op{P}_i: {\cal H}_i \rightarrow {\cal H}_i \quad i=1,2.
\end{equation}
Let $\lambda_n^{(i)}$ denote the eigenvalues of $\op{A}_i$, counted with 
multiplicity and ordered  
\begin{equation} \label{Eq:lambdai}
   \lambda_1^{(i)} \ge \lambda_2^{(i)} \ge \cdots \ge \lambda_{N_i}^{(i)}.
\end{equation}

If $\op{\rho}_i(t_0)$ is the density operator for subsystem $i$, whose matrix 
representation with respect to the basis ${\cal B}_i$ is
\begin{equation}
  \op{\rho}_i(t_0) \doteq \mbox{diag}(w_1^{(i)},\ldots,w_{N_1}^{(i)})
\end{equation}
with $w_1^{(i)} \ge w_2^{(i)} \ge \cdots \ge w_{N_1}^{(i)}$, we can apply 
theorem \ref{thm:KBI} to obtain bounds for the expectation value of $\op{A}_i$:
\begin{equation} \label{Eq:KBi}
 \sum_{n=1}^{N_i} w_{N_1-n+1}^{(i)} \lambda_n^{(i)} \le 
 \ave{\op{A}_i(t)} 
 \le \sum_{n=1}^{N_i} w_n^{(i)} \lambda_n^{(i)}. 
\end{equation}

Notice that the total probability for each subspace is less or equal to one,
and that the sum of the subspace probabilities must equal one, i.e.,
\begin{equation} 
   p_1=\sum_{n=1}^{N_i} w_n^{(i)} \le 1, \quad p_1+p_2=1.
\end{equation}
If the probability for subspace $i$ is one, then the initial ensemble is 
restricted to this subspace and since the subspaces do not interact, the 
ensemble will remain in this subspace forever, i.e., $p_i=1$ for all times.  
In this case, $\ave{\op{A}(t)}=\ave{\op{A}_i(t)}$.

If both subspaces are initially occupied, i.e., both $p_1$ and $p_2$ are 
non-zero, then the density operator for the entire space ${\cal H}$ is the direct 
sum of the subspace density operators $\op{\rho}_1(t_0)$ and 
$\op{\rho}_2(t_0)$, i.e.,
\begin{equation} \label{Eq:rho0tot}
  \op{\rho}(t_0) = \op{\rho}_1(t_0) \oplus \op{\rho}_2(t_0) 
         \doteq  \left( \begin{array}{c|c} \op{\rho}_1(t_0) & 0 \\\hline 
                                            0 & \op{\rho}_2(t_0) 
                  \end{array} \right).
\end{equation}
Since $\op{H}$ maps each subspace to itself, we can conclude
\begin{equation}
  \op{\rho}(t) = \op{\rho}_1(t) \oplus \op{\rho}_2(t) 
         \doteq  \left( \begin{array}{c|c} \op{\rho}_1(t) & 0 \\\hline 
                                            0 & \op{\rho}_2(t) 
                  \end{array} \right)
\end{equation}
for $t>t_0$ and thus it easily follows that
\begin{equation}
 \ave{\op{A}(t)} =\trace{\op{A}_1\op{\rho}_1(t)}+\trace{\op{A}_2\op{\rho}_2(t)}
\end{equation}
Since $\op{\rho}_i(t)$ and $\op{A}_i$ ($i=1,2$) are operators on $\op{H}_i$, 
we can apply (\ref{Eq:KBi}).   Thus we have 
\begin{theorem}{} \label{thm:KBII}
Consider a decoupled quantum system as defined above.  If $\op{\rho}_0$ is 
given by (\ref{Eq:rho0tot}), then the expectation value of an observable 
$\op{A}$ is bounded by
\begin{eqnarray}
 \ave{\op{A}(t)} & \ge & 
    \sum_{n=1}^{N_1} w_n^{(1)} \lambda_{N_1-n+1}^{(1)}
   +\sum_{n=1}^{N_2} w_n^{(2)} \lambda_{N_2-n+1}^{(2)} \\  
 \ave{\op{A}(t)} & \le &
    \sum_{n=1}^{N_1} w_n^{(1)} \lambda_n^{(1)}
   +\sum_{n=1}^{N_2} w_n^{(2)} \lambda_n^{(2)},
\end{eqnarray}
where $\lambda_n^{(i)}$ are the eigenvalues of the subspace observable 
$\op{A}_i$, counted with multiplicity and ordered in a decreasing sequence.
Furthermore, the upper bound is attained at $t=t_F$ if for $k=1,\cdots,m_1$ 
\begin{equation}
  \span_{j=1,\ldots,d(k)} \ket{\Psi_{r(k,j)}^{(1)}(t_F)} = E(a_k^{(1)}),
\end{equation}
where $d(k)=\dim E(a_k^{(1)})$ and $r(k,j)=d(1)+\cdots+d(k-1)+j$, and for 
$\ell=1,\cdots,m_2$
\begin{equation}
  \span_{j=1,\cdots,d(\ell)} \ket{\Psi_{r(\ell,j)}^{(2)}(t_F)} 
= E(a_\ell^{(2)}),
\end{equation}
where $d(\ell)=\dim E(a_\ell^{(2)})$ and $r(\ell,j)=d(1)+\cdots+d(\ell-1)+j$.
Similar conditions can be written down for the lower bound.
\end{theorem}
This theorem provides improved bounds for decoupled systems and it is easy
to see how it can be generalized to systems consisting of more than two 
non-interacting subsystems or control-linear systems with multiple controls.
The improved bounds are dynamically realizable if all the subsystems are 
simultaneously completely controllable.  While the previously mentioned
condition for complete controllability can be applied to each subsystem, it
is not clear whether complete controllability of all subsystems always implies 
complete controllability of the system as a whole.  However, our computations 
for several decoupled systems suggest that this is the case for the systems
we studied \cite{SGS-unpublished}.
\section{Control Computations for a Coupled Partially Controllable System}
\label{sec:coupled}
For partially controllable systems whose dynamics can not be decomposed into
independent subspace dynamics, the question of dynamical realizability of the 
kinematical bounds for a given observable can not be answered in general.  
Rather it depends on the choice of the observable.  Among the many computations 
we have done, we studied a four-level harmonic oscillator model with unusual 
interaction terms:
\[
   \op{H}_0= \left[ \begin{array}{cccc} 
                           0.5 & 0 & 0 & 0\\
                           0 & 1.5 & 0 & 0\\
                           0 & 0 & 2.5 & 0\\
                           0 & 0 & 0 & 3.5
                    \end{array} \right], \quad
   \op{H}_1=f(t) \left[ \begin{array}{cccc} 
                           0 & 1 & 0 & 0\\
                           1 & 0 & 1 & 0\\
                           0 & 1 & 0 & 1\\
                           0 & 0 & 1 & 0
                    \end{array} \right]. 
\]
The main difference of this model compared to the standard harmonic oscillator
model is that we set all the transition probabilities equal to 1 instead of 
$\sqrt{n}$.  Clearly, the system is not decoupled.  Yet, unlike the standard 
harmonic oscillator, this system is not completely controllable.  In fact, it
can be shown that the dimension of the associated Lie algebra drops from 16 
to 11 precisely when all the transition probabilities are equal.  If only one 
of these values is changed, complete controllability is recovered 
\cite{SGS-unpublished}.

Nevertheless, our control computations maximizing (a) the energy of the system, 
i.e., $\op{A}=\op{H}_0$ and (b) the transition dipole moment $\op{A}=\op{V}$, 
assuming
\[
   \op{\rho}_0 = \left[ \begin{array}{cccc} 
                           0.4 & 0 & 0 & 0\\
                           0 & 0.3 & 0 & 0\\
                           0 & 0 & 0.2 & 0\\
                           0 & 0 & 0 & 0.1
                    \end{array} \right],
\]
indicate that it still seems to be possible to control the system rather 
effectively even if all the transition probabilities are the same.  The final 
yield after 20 iterations was 97.75 \% of the kinematical maximum in case (a) 
and 95.69\% in case (b).   These results suggest that further study of the
dynamical realizability of the kinematical bounds for partially controllable
systems is necessary.
\section{Conclusion}
\label{sec:conclusion}
We have shown how the kinematical constraint of unitary evolution for 
non-dissipative quantum systems gives rise to kinematical bounds on the 
optimization of arbitrary observables and established general criteria for 
the dynamical realizability of these kinematical bounds.  It has been 
shown in particular that the kinematical bounds are always dynamically 
realizable for completely controllable systems and that improved bounds
can be derived for decoupled systems.  Finally, we have demonstrated in 
the last section that certain modifications of a completely controllable 
system may lead to a loss of complete controllability, but that even in 
such a case the kinematical bounds may still be approximately dynamically 
attainable.  In latter case, further investigation of the structure of the 
Lie algebra associated with the given control system is necessary to 
determine whether a particular kinematical bound for a partially controllable
system can be dynamically attained.
\bibliography{science}
\end{document}